\documentclass[prc,preprint,showpacs,nofootinbib]{revtex4}
\usepackage{graphicx}
\begin{document}

\title{Time-variability of alpha from realistic models of Oklo reactors}

\author{C. R.  Gould}
\affiliation{Physics Department, North Carolina State University, 
Raleigh, NC 27695-8202, USA}
\affiliation{Triangle Universities Nuclear Laboratory, Durham, 
NC 27708-0308, USA}
\email{chris_gould@ncsu.edu} 
\author{E. I.  Sharapov}
\affiliation{Joint Institute for Nuclear Research, 141980 Dubna, Moscow region, Russia}
\author{S. K. Lamoreaux }
\affiliation{Los Alamos National Laboratory, Los Alamos, NM 87545, USA}

\date{May 12, 2006}

\begin{abstract}
We reanalyze Oklo $^{149}$Sm data using realistic models of the natural nuclear reactors. Disagreements 
among recent Oklo determinations of the time evolution of $\alpha$, the electromagnetic fine structure constant, 
are shown to be due to different reactor models, which led to different neutron spectra used in the calculations.
We use known Oklo reactor epithermal spectral indices as criteria for 
selecting realistic reactor models. Two Oklo reactors, RZ2 and RZ10,  were modeled with MCNP. The resulting neutron 
spectra were used to calculate 
the change in the $^{149}$Sm effective neutron capture 
cross section as a function of a possible shift in the energy of the 97.3-meV resonance. 
We independently deduce ancient $^{149}$Sm effective 
cross sections, and use these values to set limits on the time-variation of $\alpha$.  Our study resolves 
a contradictory situation with previous Oklo $\alpha$-results.
Our suggested $2 \sigma$ bound on a possible time variation of $\alpha$ over two billion years is stringent: 
$ -0.24  \times 10^{-7} \le \frac{\Delta \alpha}{\alpha} \le 0.11  \times 10^{-7}$, but model dependent in that it assumes only $\alpha$ 
has varied over time. 
\end{abstract}
\vspace{1pc}
\maketitle

\section{Introduction}\label{sec:intro}

Two papers \cite{Lam04} and \cite{Pet05} 
on the determination of the time evolution of $\alpha$ the electromagnetic fine structure constant  from Oklo 
reactor data recently 
appeared,  adding contradictory results 
to  earlier investigations \cite{Shl76,Pet77,Dam96,Fuj00}.
The fractional change of $\alpha$ 
 over a two  billion year period has been found from Oklo data to be:
$\Delta \alpha/\alpha= +0.45^{+0.17}_{-0.07}  \times 10^{-7}$  \cite{Lam04}, 
$-0.56  \times 10^{-7}\leq \Delta \alpha/\alpha \leq 0.66  \times 10^{-7}$  \cite{Pet05},
$-0.9  \times 10^{-7} \leq \Delta \alpha/\alpha \leq 1.2  \times 10^{-7}$ \cite{Dam96}, 
and either $-0.18  \times 10^{-7} \leq \Delta \alpha/\alpha \leq 0.02  \times 10^{-7}$ or
$\Delta \alpha/\alpha= +0.88 \pm 0.07  \times 10^{-7}$ in \cite{Fuj00}. 
By comparison,   
astrophysics determinations from data on the shifts of the absorption 
lines in the spectra of quasar light have yielded 
$\Delta \alpha/\alpha = - 72^{+18}_{-18}  \times 10^{-7}$ \cite{Web01} and 
$\Delta \alpha/\alpha =  - 6^{+6}_{-6}  \times 10^{-7}$  \cite{Sri04} over an approximately ten billon year period. The sign of $\Delta \alpha$ 
is defined by the relationship $\Delta \alpha = \alpha_{past} - \alpha_{present}$, 
so that a negative sign, for example,  means that 2 - 10 billion years ago the value of $\alpha$ 
was smaller than at present.
For more results and references on the time variation of fundamental 
constants see Refs. \cite{Pei04,Uza03}.

As the results indicate, the situation is not entirely satisfactory: some analyses give only upper limits, 
while those showing a definite effect disagree even in sign. While theoretical models have been proposed which can accommodate time dependent rates of change of $\alpha$, clarifying the  disagreements among the Oklo analyses is important, particular since there are also questions about just how model-dependent these very precise limits actually are \cite{Mar84,Lan02,Cal02}.    In this paper we will concentrate on the nuclear physics aspects of the Oklo reactors, focusing 
in particular on realistic models of the neutronics.

The Oklo phenomenon has been known since 1972. The history of the discovery, the geological background, 
the relevant petrography, mineralogy, isotopic chemistry  and the Oklo reactors physics are 
definitively described by Naudet \cite{Nau91}.  
Most of details of the Oklo phenomenon  
to which we will refer are from this largely unknown text. 
%There are also   
%the IAEA proceedings of two meetings held in 1975 in Libreville \cite{Iae75} and 
%in 1978 in Paris \cite{Iae78}.   
Findings  from more recent  Oklo studies are reported  in Refs.  
\cite{Gau96} and \cite{Hid98}. \\
Sixteen  natural uranium reactors have been identified in 
Gabon, West Equatorial Africa, in three different ore deposits: 
at Oklo,  at Okelobondo 1.6 km away, and 20 km south 
of Oklo at the Bangombe. Collectively, these are called  
the Oklo Fossil Reactors. Well studied reactors include Zone two (RZ2)  
with more than sixty bore-holes,  and more recently Zone ten (RZ10) with thirteen  bore-holes.
In RZ2, 1800 kg of $^{235}$U 
underwent fission over 850 kyr of operation and in RZ10 about 650 kg of $^{235}$U 
fissioned (more rapidly) over 160 kyr of operation. All reactor zones were found  deficient in $^{235}$U, and in most of them fission products were well retained.
Isotopic concentrations were measured by mass spectrometry, and  
provided  information on the neutron fluency,  
the neutron spectral index, and the $^{235}$U restitution factor (burned $^{235}$U is partly regenerated 
after $\alpha$-decay of $^{239}$Pu formed in neutron capture on $^{238}$U). \\ 

Due to the low 0.72\%, abundance of $^{235}$U and the high np capture cross section, 
present-day natural uranium cannot sustain 
a nuclear chain reaction with light water as a moderator. However, 
2000 million years ago  \footnote{
The age of the Oklo natural reactors  is debated in the literature: Fujii et al \cite{Fuj00} and Dyson and Damour \cite{Dam96} use 2 BY. Naudet cites 1.95 BY \cite{Nau91} and 1.8 BY is used in Ref. \cite{Pet05}.},
when fission chain reactions started at Oklo,  
$^{235}$U had a relative abundance of 3.7\%, comparable to the 3$-$5\% enrichment used in most commercial power reactors.
In those times therefore a chain fission reaction was possible in principle 
and actually took place.
Reactors in the northern part of the deposit, including 
RZ2 and RZ10,  operated at a depth of several thousand meters, under then-marine sediments 
which came close to,  but still below, the surface after the tectonic uprising about 
250 millions  years ago.
 At this depth, the conditions of 
pressure and temperature are close to those of the Pressurized Water Reactors (PWR) of today 
(temperature around 300  C, pressure about 20 MPa). 
Of course, the Oklo reactor powers of 10$-$50 kW are greatly below  
the 1000-MW scale of the present reactors, and furthermore 
probably did not operate continuously.
The authors of Ref. \cite{Mes04} deduced that 
RZ13 operated for a 0.5 hour until the accumulated heat boiled away the 
water, shutting down the cycle for up to 2.5 hours until the rocks cooled sufficiently to allow water saturation to initiate a new cycle. \

Shlyakhter \cite{Shl76} was the first person to point out that a change in $\alpha$ could shift the 
position of  the 97.3-meV neutron resonance in $^{149}$Sm and that as a result the present-day capture cross section could be different from the ancient value. Assuming a reactor temperature of 300K, and taking the fission isotope abundances known at that time, he found no evidence for a shift 
in the resonance position and accordingly obtained an upper  bound for the fractional change in alpha of 0.1x10$^{-7}$ 
(a revised number from comments in Ref. \cite{Pet05}).
Using updated abundance and temperature data, Damour and Dyson 
\cite{Dam96},  
and later Fujii et al. \cite{Fuj00} carried out more detailed studies for RZ2 and RZ10.
They calculated the  present-day effective cross section 
by averaging the resonance cross section  over a presumed fully thermalized
Maxwellian neutron spectrum. 
In such an approach there is no need for a particular model for the Oklo reactors  
since the spectrum is determined solely by the temperature. 
Their results for the variation in $\alpha$ were basically in agreement, indicating no change.
By contrast, in the recent  papers \cite{Lam04} and \cite{Pet05}, where contradictory results 
have been  obtained, 
the averaging is performed over neutron spectra with 
a 1/E epithermal tail in an addition to the Maxwellian contribution. 
Spectra with different contributions from the epithermal neutron tail 
were obtained with an infinite reactor model in Ref. \cite{Lam04} 
and from  Monte Carlo modeling of a finite reactor in 
Ref. \cite{Pet05}. Not surprisingly, the use of different neutron spectra can lead 
to different results. But since these models are not unique, 
the question arises as to how to choose 
between them and between other models.\\

In the present work we suggest using 
measured  Oklo reactors epithermal spectral indices as criteria for 
selecting realistic reactor models. We perform MCNP calculations to find 
full scale models of RZ2 and RZ10 satisfying  these criteria, and we use
the resulting neutron flux spectra to calculate 
the dependence of the effective $^{149}$Sm capture cross section 
on the resonance shift. 
We deduce independently the ancient $^{149}$Sm effective neutron capture 
cross section using an updated formalism. From our limits on the 97.3-meV resonance shift, and assuming that only electroweak physics is varying, we 
we can set stringent limits on the time-variation of $\alpha$ from the Oklo data.

The paper is organized as follows: In section \ref{sec:sighat}, we review the definition of the 
effective cross section and the definitions of the various spectral indices used to define 
the contributions of the epithermal neutrons to the neutron flux. 
In section \ref{sec:model}, we present our models of the Oklo reactor zones, 
and our MCNP calculations of the neutron spectra. 
In section \ref{sec:Xaveraging} we calculate the $^{149}$Sm capture cross section 
as a function of the resonance energy shift, and in \ref{sec:Xancient} we 
review our new calculation 
of the ancient $^{149}$Sm cross section. In section  \ref{sec:results} 
we present our results and conclusions.

\section{Effective cross sections and spectral indices}\label{sec:sighat}

To analyze the Oklo reactor data without explicitly specifying
the neutron density $n(v)$ at velocity $v$,  it is customary to use not 
the average cross section $\bar{\sigma}= \int_0^{\infty}\,n(v)\,\sigma (v)\,v\,dv/\int_0^{\infty} \,n(v)\,v\,dv$, but instead an effective 
cross section defined as:
\begin{equation}
\hat{\sigma} = \int_0^{\infty}\,n(v)\,\sigma (v)\,v\,dv/nv_0.
\label{eq:sighat}
\end{equation}
Here $n = \int_0^{\infty} \,n(v)dv$ is the total density, and $v_0$=2200 m/sec is the velocity of a neutron at 
thermal energy $0.0253$ eV.
To keep $R$, the reaction rate, unchanged it is necessary also to introduce 
an effective neutron flux density $\hat{\Phi} = \int_0^{\infty} \,n(v)\,v_0\,dv$ 
different from the 'true' flux $\Phi = \int_0^{\infty} \,n(v)\,v\,dv$, leading to $R = \hat{\sigma}\,\hat{\Phi}= \bar{\sigma}\,\Phi$. 

When the cross section for a particular reaction channel exactly follows a $1/v$ law ($\sigma = \sigma_0v_0/v$), we have   
$\hat{\sigma} = \sigma_0$ and the reaction rate  
$\hat{\sigma}\hat{\Phi}=\sigma_0\, nv_0$ does not depend on the temperature $T$.
When the cross section  $\sigma$ deviates from the $1/v$ law at low energies (as it does for $^{149}$Sm), and when the neutron spectrum is not pure Maxwellian (as is the case in any realistic reactor), the effective cross section can be written 
\begin{equation}
\hat{\sigma }= g(T)\sigma_0 + r_OI, 
\label{eq:sig2}
\end{equation}
where  $g(T)$, a function of the temperature $T$, is a measure of the 
departure of $\sigma$ from the 1/$v$ law, 
$I$ is a quantity related to the resonance integral of the cross section and
$r_O$ is the Oklo reactor spectral index, a measure of the contribution of epithermal neutrons to the cross section.  The more well known epithermal
Westcott index $r$ \cite{Wes58} is related to $r_O$ by  
\begin{equation}
r_O =r\,\sqrt{\frac{T}{T_0}}
\label{eq:rft}
\end{equation}
The Westcott index  is a temperature dependent quantity while  $r_O$ (as shown below) and $I$ are independent of temperature.

Since we will be concerned with extracting $r_O$ from neutron spectra calculated by MCNP for specific reactor 
models, we follow  Ref. \cite{Wes58} and introduce 
the total neutron density $n(v)$ and its 
epithermal fraction $f_{epi}$:
\begin{equation}
n(v) = n\,\left [(1- f_{epi})\,n_{th}(v) + f_{epi}\,n_{epi}(v)\right ] ,
\label{eq:dens}
\end{equation}
where   
$n_{th}(v)=(4/\pi)(v^3/v_T^3)\,exp(-v^2/v_T^2)$ is the thermal Maxwellian distribution, and $n_{epi}(v)=v_c/v^2$ 
for $v > v_c$, otherwise zero, is an epithermal
distribution which holds for systems with zero resonance absorption.  Each of the distributions $n_{th}(v)$ and  $n_{epi}(v)$ is normalized separately to unity. The velocity 
$v_c$ is an as-yet unspecified cutoff velocity for the epithermal distribution  
and $v_T$ is the most probable neutron velocity for Maxwellian with the temperature $T$
as given by $v_T=v_0\,(T/T_0)^{\frac {1}{2}}$.
%\label{eq:vT}
The temperature $T_0$ is the temperature of the Maxwellian density distribution 
having the most probable velocity $v_0$, defined from the relation 
$mv_0^2/2 = kT_0 =0.0253$ eV.

The authors of Ref. \cite{Wes58} suggest cutting off the epithermal distribution at  energy $E_c= \mu kT$ 
with the parameter value $\mu \simeq 5$. The corresponding temperature dependent threshold velocity is then   
 $v_c=v_T\sqrt {\mu}$. At such a threshold the Maxwellian component 
 is already several times larger 
than the epithermal one, so such an approximation is satisfactory. 
The Westcott spectral index $r$ is then defined as   
\begin{equation}
r = \frac{\sqrt{\pi\mu}}{4}\,f_{epi}(T) \simeq f_{epi} .
\label{eq:r}
\end{equation} 
To assess the temperature dependence of $f_{epi}(T)$, we transform 
from neutron densities to neutron fluxes. This yields for the total thermal component
 $\phi_{th}\equiv n(1-f_{epi})\bar{v}=n(1-f_{epi})v_T\sqrt {4/\pi}$, and for an ideal 
(without resonance absorption) epithermal flux  
per unit of neutron lethargy (lethargy is $u = \ln(10/E_{\mathrm MeV})$)  
$\phi_{epi}(\Delta u=1) = n f_{epi}v_c/2$. 
Introducing the ratio 
\begin{equation}
\delta =\frac{\phi_{epi}(\Delta u=1)}{\phi_{th}},
\label{eq:delta1}
\end{equation} 
 we find:
\begin{equation}
\delta = \frac{\sqrt{\pi \mu}}{4} \frac{f_{epi}}{1-f_{epi}}.
\label{eq:delta2}
\end{equation}
For small values of $f_{epi}$ and $\mu \simeq 5$, we see that $\delta \simeq f_{epi}$. 
The quantity $\delta$ is calculated in  reactor physics \cite{Wes58} and for 
the case of moderation by hydrogen is   
\begin{equation}
\delta = \frac{\Sigma_{a}(v_0)}{ \Sigma(H)}\sqrt{\frac{\pi T_0}{4T} }.
\label{eq:delta3}
\end{equation} 
Here
$\Sigma_{a}(v_0)$ is the summed macroscopic absorption cross section 
at neutron velocity $v_0$, 
and  $\Sigma(H)$ is the macroscopic scattering cross section of  hydrogen  
in the epithermal region. 
From this equation and identifying $\delta$ with $f_{epi}$ we see that  the epithermal fraction 
of the neutron density in a reactor behaves as
$f_{epi}(T) \propto \sqrt{T_0/T}$, thereby confirming that the Oklo spectral index defined by 
Eq. \ref{eq:rft} is independent of temperature.\\

In realistic systems the epithermal flux deviates from the 1/E dependence 
due to absorption of neutrons in uranium resonances above an energy of several eV. In this case, two different definitions of $r_O$, both approximations, 
have been used. Ref. \cite{Nau91} relies on 
the shape of the neutron flux and  defines  
\begin{equation}
r_O  = \frac{\phi_{epi}(u^{\star})}{\phi_{th}^{\star}}\,\sqrt{\frac{T}{T_0}},
\label{eq:rFlux}
\end{equation} 
where $\phi_{epi}(u^{\star})$ is the is the flux per unit of lethargy at 
some effective energy in the resonance region and $\phi_{th}^{\star}$ is the 
total flux integrated up to an energy where the flux begins to increase 
above the 1/E level. In Ref. \cite{Wes58} the spectral index $r$ at room temperature is computed 
as the ratio of the effective macroscopic absorption cross section $\hat\Sigma_a$ to 
the moderating power $\bar{\xi}\Sigma_s$ \cite{Wes58}, and therefore 
\begin{equation}
r_O = \frac{\hat{\Sigma_a}}{\bar{\xi}\Sigma_s}.
\label{eq:rAbs}
\end{equation}  
We compare these three definitions in the next section. In Ref. \cite{Lam04},   the temperature dependent Weinberg-Wigner \cite{Wei58} 
thermalization parameter $\Delta (T) = 2 A \Sigma_{a}(kT)/ \Sigma_s$ 
 is used. For the case of hydrogen moderation only, it is related to $r_{O}$ 
by $r_O \approx \sqrt{\pi /4}\,\Delta (T_0)/2$. \\

The Oklo spectral indices $r_O$ are known quantities; their values   have been 
deduced for several Oklo reactor zones
\cite{Nau91}, \cite{Gau96}, \cite{Hid98}, \cite{Ruf76} from
analysis of the fission products $^{143}$Nd, $^{147}$Sm,
and the $^{235}$U concentrations.  We will use these spectral indices 
to discriminate between possible models of the ancient reactor zones.
In  particular, the following 
experimental values have been deduced: $r_O=0.20$ $-$ $0.25$ for RZ2 
and $r_{O}=0.15 \pm 0.02$  for RZ10.   The RZ2 result is a range of the borehole SC36  
values  corrected in Ref. \cite{Nau91} after re-evaluation 
of the cross section $\hat{\sigma}_{{143}_{\mathrm {Nd}}}$ to the expression 
$\hat{\sigma} = 335 - 100r_o$.  We report the RZ10 result  as 
an average  from four samples of Ref.  \cite{Hid98}.\\

\section{Oklo reactor models and neutron spectra}\label{sec:model}

The reactor criticality is determined by the size, geometry and composition of the active zone.
The latter influences the energy  dependence of the neutron flux. 
The Oklo reactor zones include uraninite ${\mathrm{UO}_2}$, gangue (oxides of different 
metals with water of crystallization), water, and  
poisons that are  present initially or build up during operation. 
Among these parameters, the most uncertain is the amount of water present at the time of reactor 
operation.
In our modeling, we vary this parameter to match the experimental spectral indices for RZ2 and RZ10 
while keeping the reactor under critical conditions. \\

The Oklo reactor cores have a characteristic horizontal size of order 10 m and occur in sandstone 
as lens shaped bodies of thickness varying between 20 and about 90 cm.
The uranium content ranged from 20 to 80 wt.\%. Each core is surrounded by a clay mantle.  
In the past
water filled spaces left by cracks and fissures. The effective porosity required to achieve 
criticality is large 
($\sim $20\%), and is explained by a de-silicication process \cite{Nau91,Gau96} consisting of  
partial leaching of 
the silica by hot thermal underground water.\\ 

\begin{table}[ht]
\caption[1]{Composition (in g/cm$^3$) and neutronic parameters of the Oklo reactors. The values shown values 
are for 2 BY ago, for example, 30 wt. \% 
of UO$_2$ in RZ10 dry ore then corresponds to  $\approx$ 22 wt. \% in  
present day ore. }\label{tab:zones}
\begin{tabular}{||c|c|c|c|c|c|c|c|c|c|c|c|c||} \hline\hline
Zone  & UO$_2$ &H$_2$O & SiO$_2$ & FeO &
Al$_2$O$_3$ & MgO  & MnO & K$_2$(Ca)O & Total & k$_{eff}$, at 300K & $r_O$ $^\dagger$\footnotetext{$^\dagger$ The $r_O$ values are calculated from the neutron densities (see text).}& $p$  \\ \hline
RZ2 &2.500 &0.636 &0.359 &0.149 &0.238 &0.077 &0.009 &0.020 &3.99 &1.033 $^\ddagger$\footnotetext{$^\ddagger$  The value for the composition with a poison of 10 ppm Boron-10 equivalent.} &0.22 &0.800\\
RZ10 &0.850 &0.355 &0.760 &0.320 &0.510 &0.160 &0.020 &0.040 &2.96 &1.036 & 0.15 &0.845\\ \hline\hline
\end{tabular}
\hspace{3mm}\\
%%%${\ddagger}$ The k$_{eff}$ value with a poison of 10 ppm Boron-10 equivalent. \\
\end{table}
We obtained the Oklo reactor neutron  
fluxes at several temperatures
with the code MCNP4C \cite{Bri00} using the free gas option for the neutron scattering.
Our model of an Oklo reactor is a flat cylinder of 70 cm height, 6 m diameter, surrounded by 
a 1 m thick reflector
 consisting of water saturated sandstone. The compositions (in g/cm$^3$) of the two reactors 
are shown in Table \ref{tab:zones}. Reactor zone RZ2 parameters are 
from Ref. \cite{Nau91}  and  
reactor zone RZ10 components are  from Refs. \cite{Gau96} and \cite{Hid98}.
 The total density of the active core material at ancient times was about 4 g/cm$^3$ for RZ2 and  
3 g/cm$^3$ for RZ10. 
The most striking difference between RZ2 and RZ10  
is  the small amount of uraninite UO$_2$ in RZ10.  In fact, with such a small amount of UO$_2$ 
it was not possible 
to make RZ10 critical with a poison more than 1 ppm of Boron-10 equivalent.  
The amount of water 
shown in Table \ref{tab:zones} is a total amount, also including water of 
crystallization.  

To check the effects of chemical bonding, we repeated MCNP calculations using the option that considers scattering from hydrogen bound 
in water molecules. The spectra were essentially unchanged except at the very lowest energies, and calculations of $\hat{\sigma}_{149}$  (see section \ref{sec:Xaveraging}) showed 1\% changes or less at all temperatures. All calculations were therefore made with the free gas option.

\begin{figure}[!ht]
\includegraphics[width=5in,angle=0]{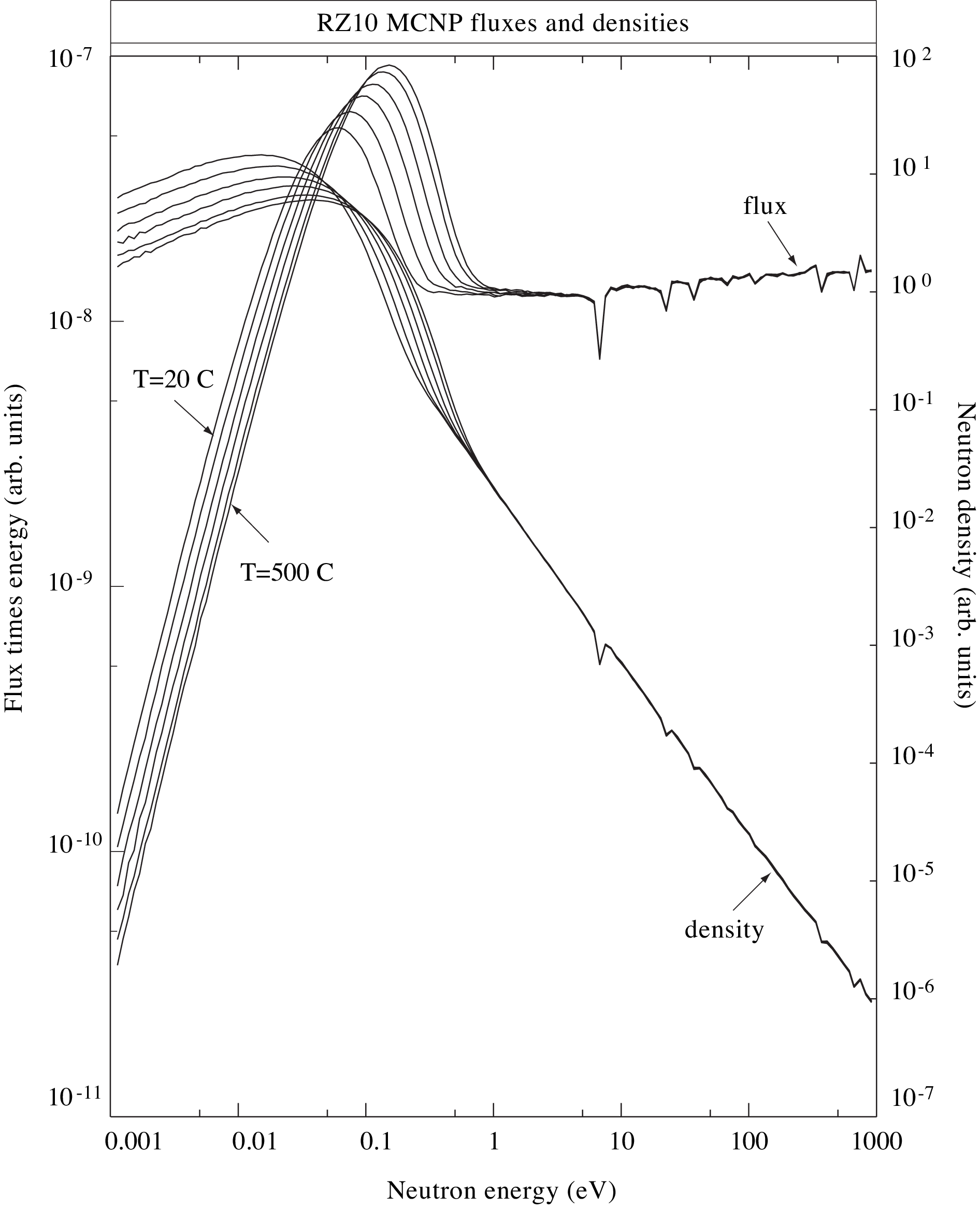}
  \caption{Reactor zone RZ10 MCNP neutron fluxes and neutron densities for
temperatures 20, 100, 200, 300, 420, 500 C. The neutron fluxes are
plotted as $\phi E$ and are the family of curves starting from a temperature 
of 20 C at the lower
left. The neutron densities
(normalized to one neutron per unit volume) are the families of curves
starting upper left.} 
\label{fig:rz10flux}
\end{figure}

\begin{figure}[!ht]
\includegraphics[width=5in,angle=0]{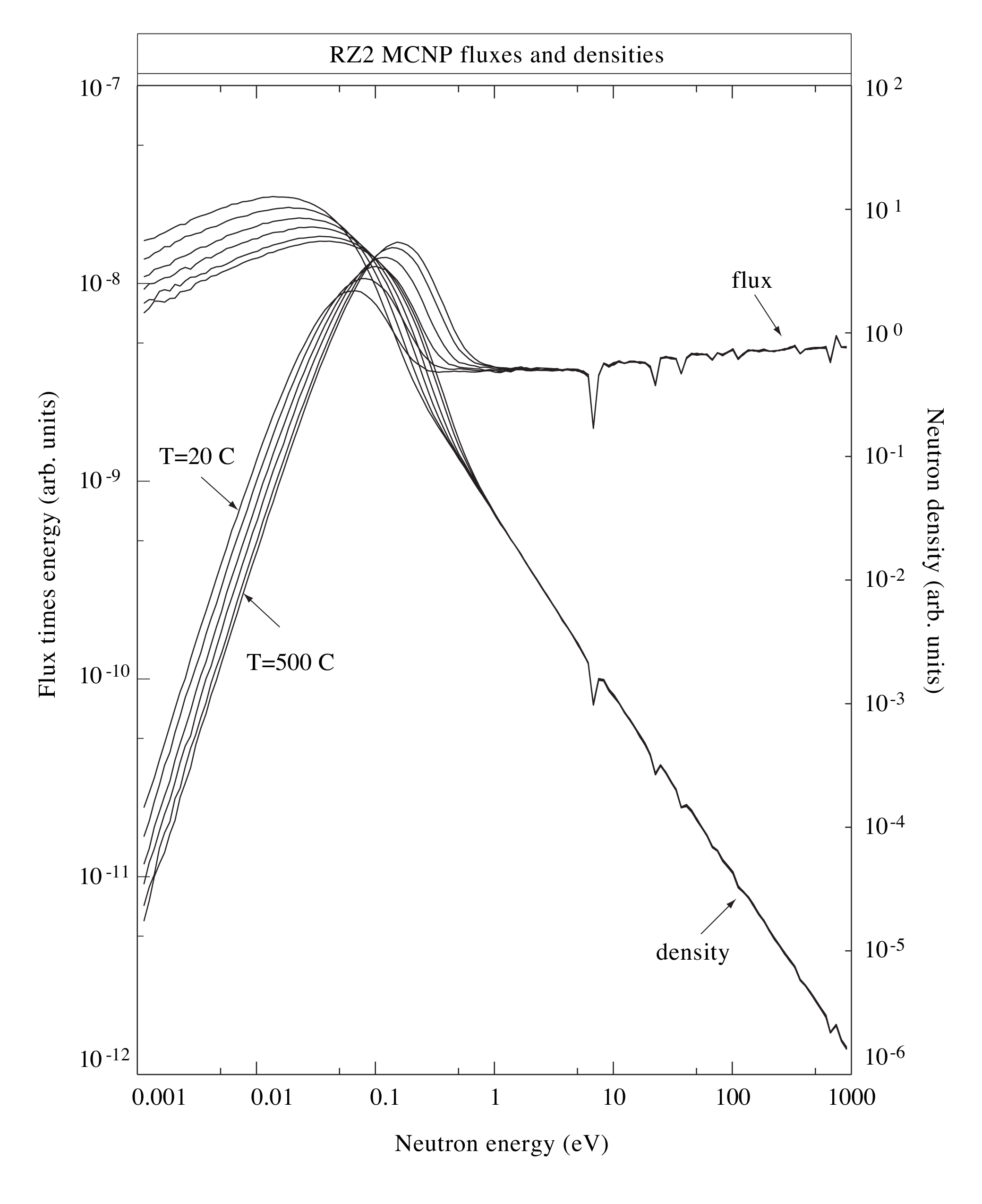}
  \caption{Reactor zone RZ2 MCNP neutron fluxes and neutron densities for
temperatures 20, 100, 200, 300, 420, 500 C. The neutron fluxes are
plotted as $\phi E$ and are the family of curves starting from a temperature 
of 20 C at the lower
left. The neutron densities
(normalized to one neutron per unit volume) are the families of curves
starting upper left.} 
\label{fig:rz2flux}
\end{figure}

The MCNP neutron fluxes and neutron densities are shown in figs \ref{fig:rz10flux} 
 and  \ref{fig:rz2flux} for temperatures
 20, 100, 200, 300, 420, 500 C. The neutron fluxes are the family of curves starting at 
the lower left of each figure. 
The leftmost curve corresponds to a temperature of 20 C and the rightmost curve to 500 C. The uranium absorption 
resonances are prominent in the epithermal region. Also of note is that the flux is not flat 
in the epithermal region, 
indicating the spectrum is not precisely 1/E. The neutron densities (normalized to one neutron 
per unit volume) are the 
families of curves starting upper left in each figure. The topmost curve corresponds to 20 
C and the lowest curve 
corresponds to 500 C. 

Oklo spectral indices can be calculated from these density plots, 
using Eqs. \ref{eq:rft} and
 \ref{eq:r}, with a cut off parameter $\mu=5$. The values we get for $r_O$ are shown in 
Table \ref{tab:zones}. 
They agree with the experimental values cited earlier, confirming that we have realistic 
models of the reactor zones. The resonance escape
probability  $p$  shown in the last column of Table \ref{tab:zones} is discussed in Section 
\ref{sec:Xancient}.

\begin{table}[ht]
\caption{Moderation characteristics of the Oklo reactor zone 
RZ2. For this reactor the hydrogen to uranium atomic ratio is $\frac{N_H}{N_U} = 7.6$, and
 the spectral index from Eq. \ref{eq:rAbs} is found to be $r_O = 0.24$.}\label{tab:rz2}
\begin{tabular} {||c|c|c|c|c|c||}\hline\hline
Atom & $\rm{N}_i, 10^{21} \rm{cm}^{-3}$ &$\Sigma_{ai}, \rm{cm}^{-1}$ &$\Sigma_{si}, \rm{cm}^{-1}$ &$\xi_i$ &$\xi_i\Sigma_{si}/\Sigma_s$\\
\hline
$^{235}$U &0.207 &0.140  &0.003 & 0.008& 0.0000\\
$^{238}$U &5.374 &0.014 &0.048 & 0.0008 & 0.0003\\
$^{10}$B  &0.011 &0.042  &0.000 & 0.210 & 0.0000\\
H  &42.60        &0.014   &0.873 & 1.000& 0.7591\\
O  &47.50        &0.0000  &0.204 &0.120 &0.0213\\
Si &3.57         &0.0006  &0.008 &0.011&0.0001\\
Fe &2.92         &0.0030 &0.003 &0.036 &0.0001\\
Al &7.30         &0.0007 &0.005 &0.070 &0.0004\\
Mg &1.35         &0.0001 &0.005 &0.080 &0.0004\\
Mn &0.08         &0.0010 &0.0003&0.036 &0.0000\\
K  &0.26         &0.0005 &0.0005&0.040 &0.0000 \\
${\mathrm {TOTALS}}$ &${\mathrm N}$ =  111.2 & $\Sigma_a$ = 0.216 & $\Sigma_s$ = 1.150 &   &$\bar\xi$ = 0.782\\
\hline
\hline
\end{tabular}
\end{table}

A specific feature of our present model as compared with Ref \cite{Lam04} 
is the presence of gangue in the reactor core. 
This means the reactor moderator is a composite substance 
with atoms besides hydrogen participating in the slowing down 
and absorption of neutrons. The contributions of these other atoms 
are presented in Table \ref{tab:rz2} for RZ2, and in Table \ref{tab:rz10} for RZ10. While only oxygen adds an 
appreciable amount to 
the moderating power  $\bar{\xi}\Sigma_s$,  (here $\bar{\xi}$ is the effective logarithmic energy 
loss defined 
as $\bar{\xi} = \sum_i \left[ \xi_i\Sigma_{si}/\Sigma_s \right]$)
several other elements contribute to the absorption parameter $\Sigma_a$. 
Using these parameters, alternative values for $r_O$ can be calculated from Eq. \ref{eq:rAbs}, 
as shown in the tables. These values agree well with the values determined from the neutron density spectra  
as shown in Table \ref{tab:zones}. We confirmed also that using the neutron flux spectra with Eq. \ref{eq:rFlux}, 
good agreement for $r_O$ values is found 
if the value for the flux per unit of lethargy is taken at about 100-eV neutron energy; 
this takes into account the deviation seen in the MCNP simulations of the flux from the 1/E law.

\begin{table}[ht]
\caption{Moderation characteristics of the Oklo reactor zone 
RZ10. For this reactor the hydrogen to uranium atomic ratio is $\frac{N_H}{N_U} = 13.0$, and
 the spectral index from Eq. \ref{eq:rAbs} is found to be $r_O = 0.14$.}\label{tab:rz10}
\begin{tabular} {||c|c|c|c|c|c||}\hline\hline
Atom & $\rm{N}_i, 10^{21} \rm{cm}^{-3}$ &$\Sigma_{ai}, \rm{cm}^{-1}$ &$\Sigma_{si}, \rm{cm}^{-1}$ &$\xi_i$ &$\xi_i\Sigma_{si}/\Sigma_s$\\
\hline
$^{235}$U &0.068 &0.0460  &0.0010 & 0.0008 &0.0000\\
$^{238}$U &1.77  &0.0048  &0.0157 & 0.0008 &0.0002\\
H         &23.76 &0.0078  &0.4871 &1.000 &0.6667\\
O         &44.39 &0.0000  &0.1910 &0.120 &0.0314\\
Si        &5.60  &0.0010  &0.0123 &0.011 &0.0002\\
Fe        &2.94  &0.0076  &0.0076 &0.036 &0.0004\\
Al        &7.29  &0.0017  &0.0017 &0.070 &0.0002\\
Mg        &3.16  &0.0002  &0.0126 &0.080 &0.0015\\
Mn        &0.18  &0.0024  &0.0004 &0.036 &0.0000\\
Ca        &0.43  &0.0002  &0.0012 &0.040 &0.0001 \\
$\mathrm{TOTALS}$ &${\mathrm N}$ = 89.59 & $\Sigma_a$ = 0.072 & $\Sigma_s$ = 0.731 &  &$\bar\xi$ = 0.701\\
\hline
\hline
\end{tabular}
\end{table}

The calculated spectral index for the infinite medium reactor model of Ref. \cite{Lam04}
 is $r_O=0.53$, which is far away from the experimental value of 0.15.
 The difference is mainly due to the small
amount of water ($\frac{N_H}{N_U} \simeq 3$) which is much less than the values we found  for RZ2, 
and RZ10. The possibility of a  larger 
amount of water in the core  was excluded  on the grounds 
that the reactor would became divergent. While this is true for an infinite reactor including uranium, 
water and a poison, it is not true for a
{\em finite}  Oklo reactor that contains an additional constituent $-$ a gangue. 
Our MCNP modeling confirmed that  
leakage of neutrons from  a finite reactor  of about 70 cm thick  
composed of only uranium and a small amount of water was greater than 
from an identical geometry reactor with gangue and much more water.
As a consequence the reactor  of the Ref. \cite{Lam04} is undercritical if it is made finite.\\

The reactor models in Ref. \cite{Pet05} are close to ours in geometry but  
the active core  compositions differ in uranium and water content.
Spectral indices $r_O$ were not reported.
The 'case 1' model of Ref. \cite{Pet05} has 1.5 g/cm$^3$ of UO$_2$ and 0.355 g/cm$^3$ H$_2$O 
at a total density
of 3.4 g/cm$^3$ with atomic ratio $\frac{N_H}{N_U} = 7.0$. This ratio is below our value of 13.0, 
therefore we believe it does not  
accurately represent RZ10. The $\frac{N_H}{N_U} = 7.0$ ratio is close to the  expected value 
for RZ2 but   the other compositions   
deviate from values established in Ref. \cite{Nau91}.\\

\section{The cross section averaging procedure}\label{sec:Xaveraging}

Following previous work, we evaluate the effective capture cross section
$\hat{\sigma}$ for $^{149}${Sm} numerically. The calculations were carried
out with the code SPEAKEASY \cite{SPE}. We include all resonances up to 51.6 eV along
with the sub-threshold resonance at $-0.285$ eV. We neglect resonance
interference terms, and use parameters from
Mughabghab et al \cite{Mug81}. In practice, the contribution of the 97.3 meV resonance dominates;
other resonances contribute only a few per cent to the sum. The bulk of
the calculations were carried out neglecting Doppler broadening; separate
calculations with the code SAMMY \cite{SAM} indicated Doppler effects on  $\hat\sigma_{149}$ were
essentially negligible for all temperatures considered,
as also found by earlier work.

The MCNP calculations provide the neutron flux per energy bin, $\phi$ on a
lethargy grid $u$ running from 23 to 9.3 in steps of 0.1. This gives
neutron energies $E = 10^7 e^{-u}$ in eV from about 1 meV up to 1 keV,
with bin widths $\Delta E = 0.1 E$. A finer grid is required for the
numerical
integrations of cross  sections for resonance neutrons since typical total resonance
widths
in $^{149}${Sm} are of order 100 meV. Accordingly, the
MCNP flux is interpolated on to a lethargy grid with step size 0.001
($\Delta E = 0.001 E$). The flux is also re-normalized by requiring the
total neutron density to sum to unity ($n=1$).

\begin{figure}[!ht]
\includegraphics[width=5in,angle=0]{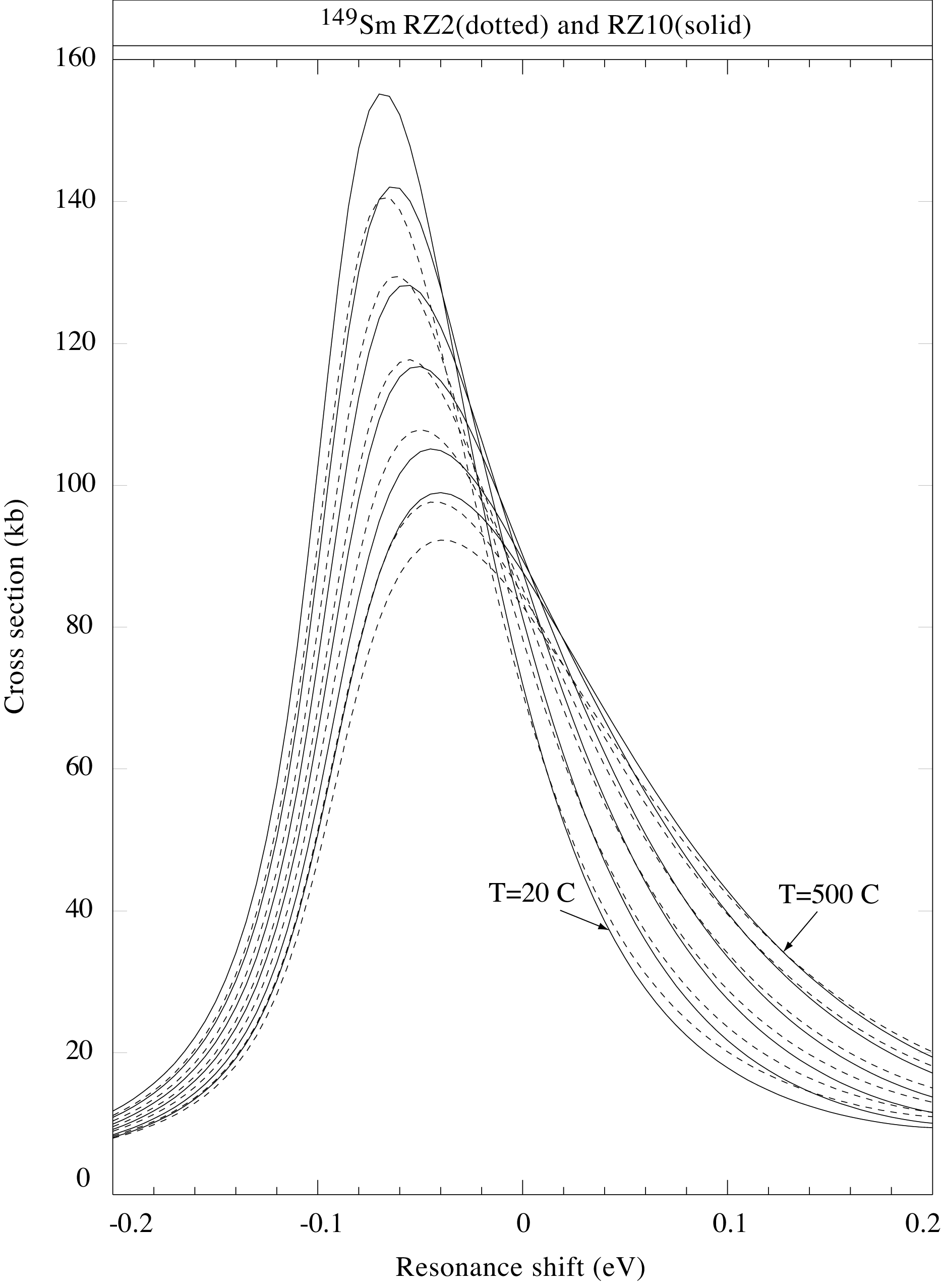}
  \caption{Calculations of the $^{149}$Sm effective capture cross section
$\hat{\sigma}_{149}$ as a function of a possible resonance energy shift
$\Delta_r$ from -200 meV to +200 meV. The results shown are for
RZ2 (dotted lines) and RZ10 (solid lines), and  for temperatures T = 20,
100, 200, 300, 420, 500 C starting from the top.} 
\label{fig:SIGMA}
\end{figure}

The effective cross section $\hat{\sigma}$ is evaluated numerically using
\begin{equation}
\hat{\sigma} = \frac{1}{v_0} \sum_i \sum_j \sigma_{ij} \phi_i.
\label{eq:e0}
\end{equation}

Here $\sigma_{ij}$ is the Breit Wigner cross section for resonance
$j$ at neutron energy $E_i$, $\phi_i$ is the neutron flux at energy $E_i$
in a bin of width $0.001 E_i$, and $v_0 = 2200$ m/s is the velocity of a
thermal neutron.

The Breit Wigner cross section $\sigma_{ij}$ for a neutron of mass $m$,
energy $E_i$ at resonance $j$ having resonance energy $E_{rj}$, neutron
width $\Gamma_{nij}$, and decay width $\Gamma_{\gamma j}$ is given by:
\begin{equation}
\sigma_{ij}=\frac{g_{0j} \pi \hbar^2}{2 m E_i}
\frac{\Gamma_{nij}\Gamma_{\gamma j}}{(E_i-E_{rj})^2+\Gamma_{ij}^2/4}.
\label{eq:e1}
\end{equation}
The energy dependent neutron width (assumed an s-wave) for resonance $j$
is given by
\begin{equation}
\Gamma_{nij}=\sqrt{E_i(eV)}\Gamma_{n0j},
\label{eq:e2}
\end{equation}
where the product of the spin statistics factor $g_{0j}$ and the reduced
neutron width $\Gamma_{n0j}$ is tabulated in Ref. \cite{Mug81}. The total
width is given by
\begin{equation}
\Gamma_{ij} = \Gamma_{\gamma j} + \Gamma_{nij}
\end{equation}
and the resonance energy is given by
\begin{equation}
E_{rj} = E_{rj0} + \Delta_r
\end{equation}

where $E_{rj0}$ is the (present day) resonance energy, and  $\Delta_r$ is the
energy shift associated with a possible change in the value of the fine
structure constant $\alpha$.

Summing over all resonances, we evaluate $\hat{\sigma}_{149}$ for values of
$\Delta_r$ from -200 meV to +200 meV. The results are shown in figure \ref{fig:SIGMA} for
RZ2 (dotted lines) and RZ10 (solid lines) for temperatures T = 20, 100, 200,
300, 420, 500 C. Temperatures run from top to bottom; in each case
T = 20 C is the top curve, and 500 C is the bottom curve.
All resonances up to 51.6 eV are included in calculations, and
are assumed to shift the same amount.
Calculations with Maxwell-Boltzman spectra were also performed; in all cases
they agreed closely with previous work.\\

\section{Calculation of the ancient capture cross section}\label{sec:Xancient}

To rederive the ancient $^{149}$Sm capture cross section, we consider the time evolution of the number densities $N_A(t)$ of the six 
isotopes of interest - $^{235}$U, $^{238}$U, $^{239}$Pu, $^{147}$Sm, $^{148}$Sm, and $^{149}$Sm. We model the time dependence of  the densities by the following set of
coupled differential equations: 
\begin{eqnarray}
{dN_5\over \hat\phi dt}&=&-\hat\sigma_{5,tot}N_5+{\lambda_9\over
\hat\phi
}N_9\\
{dN_8\over\hat\phi dt}&=&-\hat\sigma_8^0 N_8-\Gamma\\
{dN_9\over\hat\phi dt}&=&\hat\sigma_8^0 N_8 -{\lambda_9\over\hat\phi
}N_9-\hat\sigma_{9,tot}N_9+\Gamma\\
\Gamma&=&(1-p)(\nu_9\hat\sigma_{f,9}N_9+\nu_5\hat\sigma_{f,5}N_5)\\
{dN_{147}\over\hat\phi
dt}&=&-\hat\sigma_{147}N_{147}+\hat\sigma_{5,f}Y_{5,147}N_5+\hat
\sigma_{9,f}Y_{9,147}N_9\\
{dN_{148}\over\hat\phi
dt}&=&-\hat\sigma_{148}N_{148}+\hat\sigma_{147}N_{147}\\
{dN_{149}\over\hat\phi
dt}&=&-\hat\sigma_{149}N_{149}+\hat\sigma_{148}N_{148}+
\hat\sigma_{5,f}Y_{5,149}N_5+\hat\sigma_{9,f}Y_{9,149}N_9.
\end{eqnarray}
In these equations, the subscripts 5, 8, and 9 refer to $^{235}$U, $^{238}$U,
and $^{239}$Pu, respectively, and $tot$ and $f$ refer to the total
(absorption) and fission cross sections.    
 The average neutron flux, $\hat
\phi$, is assumed to be constant on a
time scale long compared to the decay constants in the problem.
The $Y$ represent the fission yields to the Sm isotopes of
interest, the subscripts 147, 148, and 149 refer to
$^{147}$Sm, $^{148}$Sm, and $^{149}$Sm, respectively, and $\nu_{5,9}$
are the average number of neutrons produced by $^{235}$U and
$^{239}$Pu, respectively.
We take the reactor start time 
as two billion years ago at $t_0=0$ and the end of operation  for RZ10 as $t_1 = 160$ kyr and for
RZ2 as $t_1= 850$ kyr. \\

Our calculation differs from the work of Fujii et al. \cite{Fuj00}
in that we explicitly incorporate the resonance escape probability, $p$ in modeling the restitution of $^{235}$U. We also include 
a term for the $^{239}$Pu alpha
decay rate, $1/\lambda_9=24.141/\ln(2)=34.8$ kyr which leads to a reduction in the restitution of $^{235}$U,  the issue being that 34.8 kyr is
not small compared to the reactor operating lifetime of 160 kyr for RZ10.

The resonance escape
probability is the probability that a neutron produced by
fission is not absorbed on the principal absorber in the system,
$^{238}$U, when moderating through the resonance region.
The quantity $(1-p)$ therefore
gives the fraction of fission neutrons that are absorbed by
$^{238}$U, most of which eventually converted to $^{239}$Pu. 
The conversion factor $C$ is defined as the ratio of atoms 
$^{239}$Pu produced to the number of atoms $^{235}$U "burned" 
in thermal neutron absorption. It is calculated 
 (Eq. (7.8) in Ref. \cite{Wei58}) as 
\begin{equation}
C = R^{-1}\frac{\hat\sigma_8^0}{\hat\sigma_{5,tot}} + (1-p)\nu_5\frac{\hat\sigma_{f,5}}{\hat\sigma_{5,tot}},
\label{eq:C}
\end{equation}
where the first term is the contribution to conversion from thermal neutrons 
(here $R$  is the fraction of atoms $^{235}$U in the  uranium fuel), 
and the second term is due to the absorption of neutrons by $^{238}$U  
in the resonance region. Not all converted atoms end  as the 
restituted $^{235}$U atoms, because some  atoms $^{239}$Pu will be burned 
in the thermal neutron flux $\hat{\phi}$.  The restitution factor $C^*$  is defined as 
\begin{equation}
C^* = \frac{C}{1+\hat{\phi}{\hat\sigma_{9,tot}\over\lambda_9}}.
\label{eq:C*}
\end{equation}

The expected value of $p$ is shown in Table \ref{tab:zones} and is calculated  as follows. 
 For U in 
a composite moderator, the definition of the resonance escape probability is 
\begin{equation}
p=\exp\left(-{1\over\bar{\xi} \Sigma_s}\int{\Sigma_a\over 1+{\Sigma_a/\Sigma_s}}{dE\over E} \right) \equiv \exp\left[- \frac{{\rm N}_{\rm U}}{\bar{\xi} \Sigma_s} I_{eff}^{res}  \right],
\label{eq:p}
\end{equation}
where the effective resonance integral $I_{eff}^{res}$ for a homogeneous 
mixture of fuel and moderator  was found in experimental and theoretical works \cite{Wei58}  (Eq. (10.29)) to be
\begin{equation}
I_{eff}^{res} = 3.8 \left( \frac{10^{24}\Sigma_s}{{\rm N}_{\rm U}} \right)^{0.42}.
\label{eq:Ieff}
\end{equation}
In these equations, 
$\Sigma_s$ is the total macroscopic 
scattering cross section of all elements in the active core, and $\bar {\xi}$ is the 
corresponding effective logarithmic energy loss defined by (Ref. \cite{Wei58}, Eq. (10.20)) 
as $\bar{\xi} = \sum_i \xi_i\Sigma_{si}/\sum_i\Sigma_{si}$.

Note that these equations assume that the atoms are free for the energy
loss, i.e., chemical bond effects are neglected.  Because the
resonance integral is over the range 0.5 eV to 100 keV, the
effects of chemical bonds are expected to be small. 
From these equations and using the RZ10 characteristics from Table \ref{tab:rz10}, 
we calculate for RZ10 the value $(1-p) = 0.155$. Eq. \ref{eq:C} gives $C(RZ10) = 0.11 + 2.05(1-p) = 0.43$. For the {\it metasample} RZ10  flux (see later) we then get $C^*=0.77C = 0.33$,  in reasonable agreement with $C^* = 0.38$, the average of the values quoted in Hidaka and Holliger (HH) \cite{Hid98}. 
\\

The coupled equations are solved with input parameters \cite{Nau91} listed in (Table \ref{tab:sigmas}). We confirmed the correctness of these effective cross sections by also calculating 
the integral $\hat\sigma=\int \sigma(E)\Phi(E,T)dE/ \int \Phi(E,T)/
\sqrt{E/0.0253}dE$) with our MCNP fluxes. Because the resonance escape
probability represents the capture probability for $E>0.5$ eV,
$\hat\sigma_8^0$ is determined by integrating to 0.5 eV, the rest of
the resonance integral being incorporated into the resonance
escape probability.

\begin{table}[ht]
\caption[1]{Effective cross sections for RZ10 and RZ2 used to determine
$\hat\sigma_{149}$ (Cross sections are based on data from Naudet. Numbers in the first row are
for the zone RZ10  and in the second row - for RZ2).}
\label{tab:sigmas}

\begin{tabular}{||c|c|c|c|c|c|c|c|c|c|c|c|c||} \hline\hline
$\hat\sigma_{5}$, kb &$\hat\sigma_{5,f}$, kb &$\hat\sigma_{9}$, kb& $\hat\sigma_{9,f}$, 
kb &$\hat\sigma_{8}^0$, kb &$\hat\sigma_{147}$, kb &$\hat\sigma_{148}$, kb& $Y_{5,147}$ 
& $Y_{9,147}$ &$Y_{5,149}$ &$Y_{9,149}$ &$\nu_5$ &$\nu_9$\\ 
\hline
0.656 &0.549 &2.05 &1.35 &0.0027 &0.142 &0.0024 &0.0226 &0.0226 &0.011 &0.011 &2.43 &2.88\\
0.665 &0.550 &2.16 &1.40 &0.0027 &0.184 &0.0024 &0.0226 &0.0226 &0.011 &0.011 &2.43 &2.88\\
\hline
\hline
\end{tabular}
\end{table}

The coupled equations are solved as follows. First, fixing $1-p = 0.155$ for RZ10, and with starting parameters $N_5(0)=0.0370$, $N_8(0)=0.963$,
$N_9(0)=0$, we can solve the first three equations for the fluence $\hat{\phi} t_1$ which reproduces the $N_5(t_1)/N_8(t_1)\equiv (N_5(t_1)+N_9(t_1))/N_8(t_1)$ ratio
determined by the present-day HH data for each RZ10 sample. We use $T_{O} = 2 \times 10^{9}$ years ago to derive the ratios at time $t_1$. We also calculate values for a {\it metasample} which represents an average of the
individual sample properties. The idea of the metasample is that there could
have been internal mixing between the samples or other processes. It
is also useful to see whether the sample averages provide
results in a reasonable range.

With the fluence fixed, we next vary the starting amount of elemental Sm relative to U,
$N_{\rm Sm}(0)/N_{\rm U}(0)$ to reproduce the ending isotopic fraction $f_{147}(t_1)$, and then vary the cross section $\hat{\sigma}_{149}$ to reproduce the ending isotopic fraction $f_{149}(t_1)$.  
We use the natural abundancies to convert from an elemental starting ratio to the isotopic starting ratios ($f_{144}(0)=0.031$,$f_{147}(0)=0.151$, $f_{148}(0)=0.113$, and
$f_{149}(0)=0.139$). In principle the starting fraction of Sm relative to U, while very small, is not a free parameter; it is determined by data for $f_{144}(t_1)$ since $^{144}$Sm is not produced in fission: 
\begin{equation}
f_{144}(t_1)N_{\rm Sm}(t_1)=f_{144}(0)N_{\rm Sm}(0).
\end{equation}

In practice, while $\hat{\sigma}_{149}$ can be tuned closely to match $f_{149}(t_1)$ we were not able to reproduce $f_{147}(t_1)$ without also varying $N_{\rm Sm}(0)/N_{\rm U}(0)$. 
Table \ref{tab:results} summarizes our results. The measured values from HH are shown above the empty line and our calculated values for $N_5(t_1)/N_8(t_1)$, $f_{147}(t_1)$ and $f_{149}(t_1)$ are shown below the empty line, along with our calculated fluences, calculated starting fractions of elemental Sm relative to U, and calculated $\hat{\sigma}_{149}$ values for the four RZ10 samples and the metasample. The mean of the four values is $89.4$ kb, with sample standard deviation $6.8$ kb.

The starting fractions, $N_{\rm Sm}(0)/N_{\rm U}(0)$ vary, but are consistent with HH data for bore holes 1640 and 1700 outside the reactor zones, and apart from 1480 are within a factor of two of values predicted by the HH data. The mean value for the ending amount of $N_{\rm Sm}(t_1) = 74.6 \pm 19.6$ $\mu$gms/gm agrees well with the mean of the measured HH values $67.8 \pm 36.0$ $\mu$gms/gm, also giving us confidence that the starting ratios are reasonable. 

As a final value for RZ10 we adopt the cross section extracted for the metasample, and to take account of sample-to-sample variations we adopt the sample standard deviation as a $1 \sigma$ uncertainty:
\begin{equation}
\hat{\sigma}_{149} = 85.0 \pm 6.8 {\rm kb.}
\end{equation}
This is in good agreement with the mean of the four RZ10 samples analyzed by Fujii et al. \cite{Fuj00}: $ 91.2 \pm 7.6 $ kb, again taking their sample standard deviation as a $1 \sigma$ error.  

Data for fifteen RZ2 samples were analyzed by Damour and Dyson \cite{Dam96}, who set a $2 \sigma$ bound  
$ 57 \le \hat{\sigma}_{149} \le 93 $ kb. Results from RZ2 have tended to show more scatter because the samples come from mining near the surface with a greater potential for contamination compared to the deep underground samples for RZ10. Nevertheless we repeated our calculations for the five RZ2 samples from borehole SC36 - 1408 to 1418 - cited by Ruffenach \cite{Ruf78} as having the most important contributions of elements formed in fission.  We followed the same procedure as for RZ10, also including analysis of a metasample formed by averaging the input data for the five individual samples. With ($1-p$) fixed, we varied the flux to match the U isotope ratios, and then varied $\hat{\sigma}_{149}$ and the starting elemental ratio of Sm and U to reproduce  the ending $f_{147}(t_1)$ and $f_{149}(t_1)$ fractions. Table \ref{tab:rz2data} summarizes the results.  The average of the five values is 75.6 kb with a sample standard deviation of 10.0 kb. The metasample value is 71.5 kb, and we therefore adopt as our RZ2 result:

\begin{equation}
\hat{\sigma}_{149} = 71.5 \pm 10.0 {\rm kb,}
\end{equation}
a value consistent with the RZ10 result and with the Damour and Dyson analysis.    

 \begin{table}[ht]

\caption{RZ10 calculated fluences $\hat\phi t_1 $ and ancient cross sections $\hat\sigma_{149}$. Input data for Sm and U isotope fractions (above the empty line) are from Ref. \cite{Hid98}. Calculated values are below the empty line. For these calculations $t_1= 160$ kyr. The metasample input data are an average of the data for the four individual samples.}

\label{tab:results}

\begin{tabular}{||c|c|c|c|c|c||}\hline
Property&SF84-1469&SF84-1480&SF84-1485&SF84-1492&Meta\\ \hline
${N_{5}(t_1)\over N_{8}(t_1)}$ &0.03176&0.02662&0.02967&0.03042&0.02962\\
$f_{144}(t_1)$ &0.001052&0.002401&0.002073&0.001619&0.001786 \\
$f_{147}(t_1)$ &0.5534&0.5323&0.5403&0.5481& 0.5435\\
$f_{149}(t_1)$ &0.005544&0.002821&0.004466&0.004296&0.004281  \\
\hline
$\hat\phi t_1\ {\rm (kb)^{-1}} $ &0.475&0.915&0.645&0.585&0.650 \\
${N_{5}+N_{9}(t_1)\over N_{8}(t_1)}$ &0.03177&0.02666&0.02969&0.03041&0.02963\\

$N_{\rm Sm}(0)\over N_{\rm U}(0) $ &12.0 $\times 10^{-6}$&50.5 $\times 10^{-6}$&32.1 $\times 10^{-6}$&22.6 $\times 10^{-6}$&27.6 $\times 10^{-6}$ \\
$f_{147}(t_1)$ &0.5532&0.5323&0.5404&0.5483& 0.5434\\
$f_{149}(t_1)$ &0.005542&0.002836&0.004470&0.004289&0.004283  \\
$\hat\sigma_{149}$ (kb) &94&86&81.5&96&85 \\
\hline
\end{tabular}

\end{table}

\begin{table}[ht]
\caption[1]{RZ2 borehole SC36 calculated fluences $\hat\phi t_1 $ (column 4) and ancient cross sections $\hat\sigma_{149}$ (column 8). Input data for Sm and U isotope fractions (columns 2,5,6 and 7 are from Ref. \cite{Ruf78}. For these calculations $t_1 = 850$ kyr. The metasample input data are an average of the data for the five individual samples.}
\label{tab:rz2data}
\begin{tabular}{||c|c|c|c|c|c|c|c|c||} \hline\hline
Sample &${N_{5}(today)\over N_{8}(today)}$& ${N_{5}(t_1)\over N_{8}(t_1)}$ &$\hat\phi t_1\, {\rm (kb)^{-1}}$  & 
$f_{144}(t_1)$ & $f_{147}(t_1)$ & $f_{149}(t_1)$&  $\hat\sigma_{149}$ (kb)  \\ \hline
1408 &0.00545 & 0.02864 & 0.905   &0.0046 &0.506 &0.0036 & 69  \\ 
1410 &0.00527 & 0.02767 & 1.015   &0.0015 &0.532 &0.0030 & 78  \\
1413 &0.00410 & 0.02153 & 1.820   &0.0007 &0.534 &0.0018 & 65  \\
1416 &0.00522 & 0.02741 & 1.045   &0.0003 &0.534 &0.0024 & 91  \\
1418 &0.00574 & 0.03013 & 0.745   &0.0003 &0.541 &0.0042 & 75  \\
\hline
meta &0.00516 & 0.02708 & 1.085   &0.0015 &0.529 &0.0030 & 71.5  \\
\hline
\hline
\end{tabular}
\end{table}

\section{Results for resonance shift and change in $\alpha$}\label{sec:results}

To calculate bounds on the possible shift of the $^{149}$Sm resonances over time we use the RZ10 result $\hat{\sigma}_{149} = 85.0 \pm 6.8$ kb, and conservatively adopt  a $ \pm 2 \sigma$ range to establish limits: $ 71.4 \le \hat{\sigma}_{149} \le 98.6 $ kb. We further assume that the reactor operated between 200
and 300 C as per the analysis by Meshick et al. \cite{Mes04}, which limits the
temperature to this range, and as per knowledge that the geological formation and depth
would not allow liquid water to be present above 300 C. 

\begin{figure}[!ht]
\includegraphics[width=5in,angle=0]{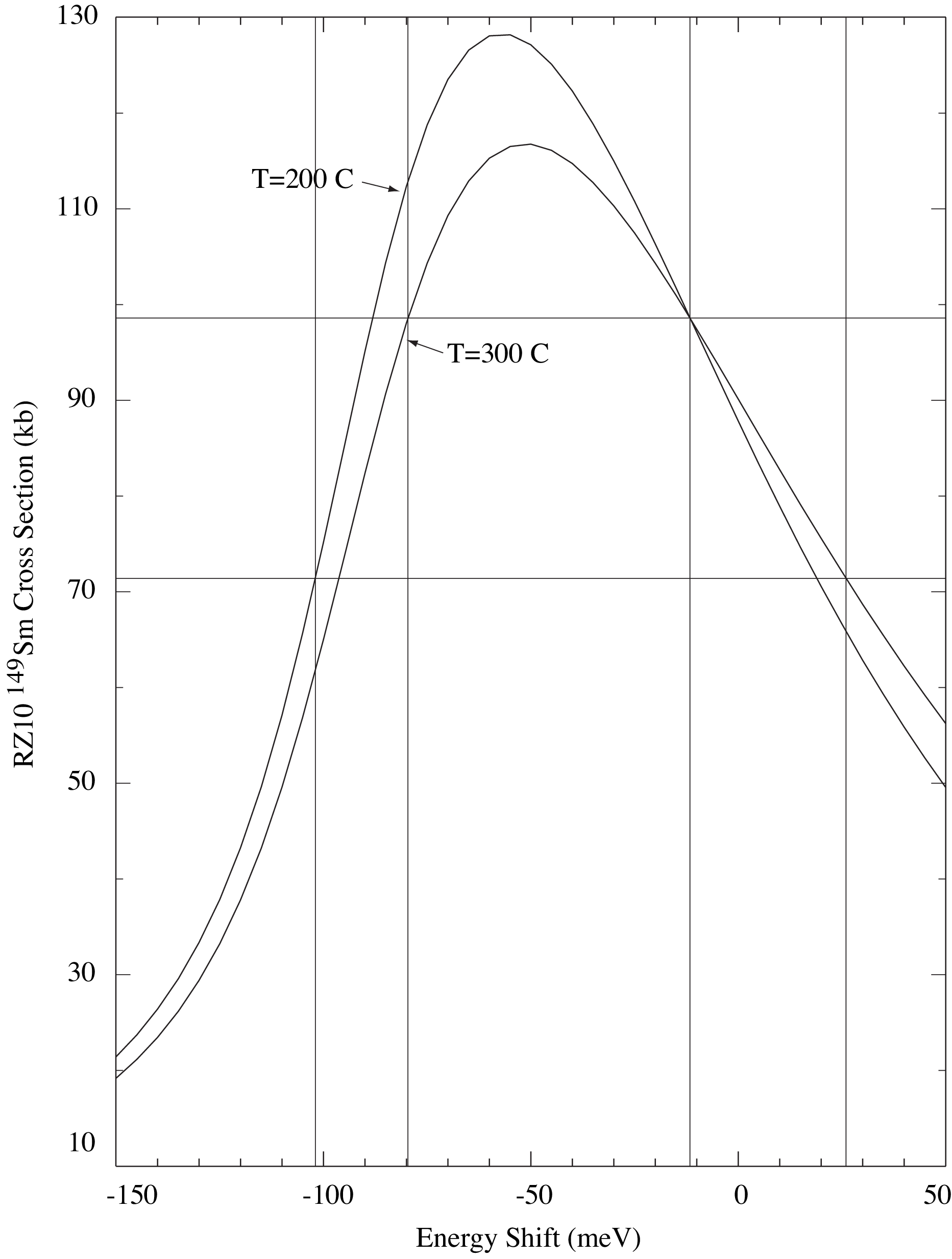}
  \caption{Expanded version of the RZ10 cross section plot of fig. \ref{fig:SIGMA} for temperatures 200 C and 300 C, and the bounds on $\Delta_r$ obtained from our $\pm 2 \sigma$ limits on the $\hat{\sigma}_{149}$.}  
\label{fig:rz10}
\end{figure}

An expanded version of the calculated RZ10 $\hat{\sigma}_{149}$ cross sections at 200 and 300 C is shown in fig. \ref{fig:rz10}. From it we obtain two solutions for the resonance energy shift $\Delta_r$:
\begin{equation}
-11.6 {\rm meV} \le \Delta_r \le +26.0 {\rm meV}
\end {equation}
\begin{equation}
-101.9 {\rm meV} \le \Delta_r \le -79.6 {\rm meV}
\end {equation}
The left-branch negative solution is tentatively ruled out by Fujii et al. \cite{Fuj00} on the basis of other data, but we retain it here for completeness.

Damour and Dyson \cite{Dam96} analyzed the dependence on $\alpha$ of the Coulomb energy contribution 
to the resonance  energy of $^{149}$Sm 
 and concluded that an energy shift due to the electromagnetic 
interaction was related to a shift in $\alpha$ by 
\begin{equation}
\frac{\Delta \alpha}{\alpha} = - \frac{\Delta_r }{1.1\;{\mathrm MeV}}. 
\label{eq:damour}
\end{equation}
The energy ${1.1\pm 0.1}$ MeV was estimated from known  isotopic shifts 
in the mean square radii of the proton density distributions in samarium.

With our results for the bounds on the 97.3-meV resonance shift $\Delta_r$ in $^{149}$Sm 
in Eq. \ref{eq:damour} we therefore obtain $2 \sigma$ bounds on the $\alpha$-change of:
\begin{equation}
-0.24 \times 10^{-7} \le \frac{\Delta \alpha}{\alpha} \le 0.11  \times 10^{-7} 
\label{eq:result}
\end{equation}
or 
\begin{equation}
0.72  \times 10^{-7} \le \frac{\Delta \alpha}{\alpha} \le 0.93  \times 10^{-7}.
\label{eq:result1}
\end{equation}

Our result is quite insensitive to the  assumption of fixed $p$.  Varying the H/U ratio changes $p$, and therefore affects the neutron spectrum and 
the resulting effective cross section.  The major effect is due to change in the overlap
of the neutron spectrum with the $^{149}$Sm 97.3-meV resonance.  Combining the
effect on the derived $\hat{\sigma}_{149}$ and the change in the overlap
of the neutron spectrum, we find $\delta(\Delta_r) \sim (50 {\rm meV}) \cdot \delta p$. A 10\% change in $p$ therefore leads to a change in $\Delta_r$ of order 5 meV, well within the $2 \sigma$ bounds assumed in extracting our final result.

In their analysis Damour and Dyson assumed that 
changes in the nuclear part of the Hamiltonian did not correlate with 
with changes in the Coulomb energy over time. They noted that 
changes in the $m_q/m_p$ could, in principle, show themselves as changes of the 
resonance position. However, at that time there was no theory to allow estimates of such effects.
Recent QCD-based developments have speculated that a time variation in $\alpha$ 
could be accompanied by a larger (up to a factor of 30 or more) variation 
of the QCD scale parameter $\Lambda$ (=~ 213 MeV) \cite{Lan02,Cal02, Fla03}
which characterizes the masses of participating particles.
Changes in the effective nuclear potential 
from, e.g. changes in $m_{\pi}/m_p$ could, in principle, therefore have a significant effect 
on the resonance shift. A detailed theoretical analysis and quantitative estimates  
of the nuclear physics aspects of the neutron resonance shift remain to be carried out. In such a situation the present result from the Oklo reactor data while stringent - and consistent with no shift in $\alpha$ over a two billion year period - should 
be regarded as a model dependent \cite{Pei04,Mar84}.

\acknowledgments
We would like to thank Dr. R. Golub for fruitful discussions 
and  Prof. Raymond L. Murray for comments related to the nuclear reactor physics.\\

This work was supported by the US Department of Energy,
Office of Nuclear Physics, under Grant No.
DE-FG02-97ER41041 (NC State University) and Grant LANL LDRD 20040040DR (Los Alamos National Laboratory)

\end{document}